\documentclass[11pt,aps,superscriptaddress,nofootinbib,preprint]{revtex4}
\usepackage{graphics}
\usepackage{graphicx}
\usepackage{epsfig}
\usepackage{psfrag}
\usepackage{color}

\usepackage{amsfonts}
\usepackage{amssymb}

\newcommand*{\be}{\begin{equation}}
\newcommand*{\ee}{\end{equation}}
\newcommand*{\bea}{\begin{eqnarray}}
\newcommand*{\eea}{\end{eqnarray}}

\providecommand*{\ler}{\stackrel{\scriptstyle <}{\scriptstyle \sim}}
\providecommand*{\ger}{\stackrel{\scriptstyle >}{\scriptstyle \sim}}

\newcommand{\eref}[1]{(\ref{#1})}

\newcommand{\comment}[1]{}

\begin{document}
\title{Running $U_{e3}$ and  BR($\mu\to e +\gamma$) in SUSY-GUTs }
\author{L. Calibbi}
\email[E-mail: ]{calibbi@pd.infn.it}
\affiliation{ Dipartimento di Fisica `G. Galilei' and
INFN, Sezione di Padova,
Universit\`a di Padova, Via Marzolo 8, I-35131, Padova, Italy.}
\author{A. Faccia}
\email[E-mail: ]{faccia@pd.infn.it}
\affiliation{ Dipartimento di Fisica `G. Galilei' and
INFN, Sezione di Padova,
Universit\`a di Padova, Via Marzolo 8, I-35131, Padova, Italy.}
\author{A. Masiero}
\email[E-mail: ]{masiero@pd.infn.it}
\affiliation{ Dipartimento di Fisica `G. Galilei' and
INFN, Sezione di Padova,
Universit\`a di Padova, Via Marzolo 8, I-35131, Padova, Italy.}
\author{S. K. Vempati}
\email[E-mail: ]{vempati@cpht.polytechnique.fr}
\affiliation{Centre de Physique Theorique
\footnote{Unit{\'e} mixte du CNRS et de l'EP, UMR 7644.},
Ecole Polytechnique-CPHT, 91128 Palaiseau Cedex, France}
\affiliation{The Institute of Mathematical Sciences, Chennai 600 013, India}

\begin{abstract}
In supersymmetric seesaw models based on SUSY-GUTs, it could
happen that the neutrino PMNS mixing angles are related to
the lepton flavour violating decay rates. In particular
SO(10) frameworks, the smallest mixing angle would get directly
correlated with the $\mu \to e + \gamma $ decay amplitude.
Here we study this correlation in detail considering $U_{e3}$
as a free parameter between 0 and $U_{e3}$(CHOOZ). Large radiative
corrections to $U_{e3}$ present in these models, typically of the
order ~$\Delta U_{e3} ~\sim ~10^{-3}$ (peculiar to hierarchial
neutrinos), can play a major role in enhancing the Br($\mu \to e
+\gamma$), especially when $U_{e3} \ler 10^{-3}$. For large
tan$\beta$, even such small enhancements are sufficient to bring the
associated Br($\mu \to e + \gamma$) into realm of MEG experiment as
long as SUSY spectrum lies within the range probed by LHC. On the
other hand, for some (negative) values of $U_{e3}$, suppressions
can occur in the branching ratio, due to cancellations among different
contributions. From a top-down perspective such low values of
$U_{e3}$ at the weak scale might require some partial/full cancellations
between the high scale parameters of the model and the radiative
corrections unless $U_{e3}$ is purely of radiative origin at the high scale.
We further emphasize that in Grand Unified theories there
exist additional LFV effects related to the running above the GUT scale,
that are also independent on the low energy value of $U_{e3}$.
These new contributions can become competitive and even dominant
in some regions of the parameter space.
\end{abstract}
\maketitle

\section{Introduction and Conclusions}
It is well know that, after the discovery that neutrinos are massive, the
detection of supersymmetry (SUSY) induced lepton flavour violation
(LFV) processes has become a very interesting possibility. This is specifically
true in the presence of a seesaw like mechanism being operative at the
high scale leading to small non-zero neutrino masses as well as large
mixing in the neutrino sector. The potential of this admixture of SUSY
and seesaw \cite{amfb} and its implications to lepton flavour violation
has been studied in several papers in the last few years
\cite{recentlfv}, especially in the light of upcoming experiments
like MEG \cite{review}.

Thanks to the RG evolution (in the presence of heavy right handed neutrinos)
from the high scale to the low energies where experiments are conducted,
SUSY seesaw leads to potentially sizable mixing
effects in the slepton mass matrices, which give rise
to flavour violating charged leptonic decays through loop-induced
processes. These flavour violating effects are strongly dependent
on the neutrino Yukawa matrix, $Y_\nu$ whose entries are generically
unknown. Fortunately, the seesaw mechanism fits nicely within the larger
picture of SUSY Grand Unification (GUTs) especially in models based on
$SO(10)$ gauge groups. A quite general feature of SUSY $SO(10)$ is the relation
among the $Y_\nu$ and up-quark Yukawa matrix $Y_u$ eigenvalues, which
ensures that at least one of the neutrino Yukawa is as large as the
top Yukawa $y_t$ \cite{oscar}.

In previous works \cite{oscar,calibbi}, we considered two benchmark
cases in which the mixing angles in $Y_\nu$ were either minimal
(CKM-like) or maximal (PMNS-like). In those works, we have studied
the implications of these two benchmark scenarios for the (indirect)
discovery potential of SUSY in the various up-coming experiments
like MEG, PRISM/PRIME, and super-B factories. We have found that
experimental sensitivity of these experiments can be quite
complementary with the direct discovery machine, Large Hadron
Collider (LHC) at CERN and in some cases, could far outreach the
sensitivity of LHC itself. These latter cases are the ones which
have large (maximal) mixing in the neutrino Yukawa couplings going
by the name, PMNS-case.  A particular feature of this case, is that
some of the LFV processes, such as $\mu \to e\; \gamma$, turn out to
be dependent on the so far unknown $U_{e3}$ entry of the mixing
matrix $U_{\mathrm{PMNS}}$. The aim of the present paper is to study
the correlation between the unknown entry $U_{e3}$ of the PMNS
mixing matrix of neutrinos, and the LFV decay, $\mu \to e\; \gamma$,
in a $SO(10)$ scenario with mSUGRA boundary conditions. To this
extent, we will treat the low energy value of $U_{e3}$ as an
independent parameter varying between $0 \ler U_{e3} \ler
U_{e3}({\rm CHOOZ})$ \cite{chooz}.

In studying the variation of the BR($\mu \to e + \gamma$) with respect
to the unknown neutrino mixing angle $U_{e3}$ a crucial factor turns out
to be the RG effects on the neutrino mass matrices
themselves from the weak scale to the scale of right handed neutrinos
and further up to the GUT scale (in terms of the seesaw parameters).
The main point is that even small $U_{e3}$ can be sufficient to generate
large corrections to the BR, thanks to running effects of $U_{e3}$
itself. In fact, the crucial parameter in computing BR($\mu \to e + \gamma$)
turns out to be the high-energy value of $U_{e3}$, instead the low-energy
one, which can be measured by neutrino oscillations experiments.
Even small values of the $U_{e3}$
generated at the high scale $\sim~ \mathcal{O}(10^{-3}-10^{-4})$ could
significantly modify the RG generated flavour off-diagonal entries
in the slepton mass matrices and thus enhancing the branching
ratios. Perhaps, the most striking aspect of this appears in the
predictions of the branching ratios for $\mu \to e + \gamma$ at the
MEG experiment in the SUSY-GUT parameter space being probed by LHC.
In fact, including these effects would enhance the predictions for the branching
ratios by an order to a couple of orders of magnitude, and thus predicting
a positive signature for $\mu \to e + \gamma$,
at least in the large $\tan\beta$ regime.
This particular aspect has been already pointed in passing in our previous
work \cite{calibbi}. More recently, the correlation between $U_{e3}$ and other
low energy observables in a purely bottom-up approach has been the subject
of a thorough analysis of Antusch et al.  \cite{herrero}.
 We'll comment more about the complementarity of their analysis
with our present study. Finally, we note that we resort to purely
phenomenological approach without worrying about aspects of flavour
model building and origins of $U_{e3}$ and other mixing angles in this
work. Such an interesting and important analysis will be treated elsewhere.

In the present work, we study the implications of \textit{running} $U_{e3}$
within the context of SUSY-GUTs.
An important point to emphasize is that
in SUSY-GUTs there exist other LFV contributions, which rely on the running
from the superlarge scale of supergravity breaking down to the GUT scale
and are independent of $U_{e3}$. This paper intends to study the interplay of the
two above mentioned sources of LFV for the observability of $\mu \to e + \gamma$.
The two sources will be discussed and compared in Sect. II and III. The main
conclusions will be drawn in Sect. IV.

\section{Running $U_{e3}$ and $(\Delta_{LL})_{12}$}
The correlation between $U_{e3}$ and $\mu \to e + \gamma$ within the
context of SUSY seesaw has been pointed out long ago \cite{tobeyana} and
further reviewed by various authors in the recent times \cite{review,herrero}.
In SUSY seesaw models, the crucial point is the strong dependence of LFV
effects on the unknown $Y_{\nu}$ matrix. Such uncertainty is due to
the fact that the high-energy parameters entering the seesaw mechanism
($Y_{\nu}$, $M_R$) cannot be obtained in terms of the low-energy neutrino
masses and mixings ($m_{\nu_k}$, $U_{PMNS}$), simply because the number of
the high-energy parameters is larger. Even within a general SUSY $SO(10)$
framework where the eigenvalues of $Y_{\nu}$ are related to the up-quark
Yukawas, it is necessary to make assumptions about the mixing structure
of $Y_{\nu}$. An interesting possibility is the case in which the mixing
angles result to be PMNS-like. This is what we called PMNS (maximal) mixing
case \cite{oscar,calibbi}. Here the `left'-mixing present in the neutrino
Yukawa matrix follows the neutrino mixing matrix and is given as :
\begin{equation}
Y_{\nu} = U_{\mathrm{PMNS}} Y_u^{diag},
\label{PMNS}
\end{equation}
The boundary condition given in Eq. \eref{PMNS} can be, for instance,
achieved starting
from the $SO(10)$ superpotential \cite{Moroi:2000tk}:
\be
W_{SO(10)} = (Y_u)_{ij} {\bf 16}_i {\bf 16}_j {\bf 10}_u +
               (Y_d)_{ii} {\bf 16}_i {\bf 16}_i
               {\langle{\bf 45}\rangle \over M_{\mathrm{Planck}}} {\bf 10}_d
               + (Y_R)_{ij} {\bf 16}_i {\bf 16}_j {\bf 126}
\label{superpotential}
\ee
where ${\bf 16}$ is the $SO(10)$ matter representation ($i,j$ are flavour
indices), ${\bf 10}_u$, ${\bf 10}_d$, ${\bf 45}$ and ${\bf 126}$ are Higgs
representations and $Y_u$, $Y_d$ and $Y_R$ are Yukawa couplings. The term with
${\bf 126}$ uniquely gives rise to right-handed neutrinos masses.

In the case Eq. \eref{PMNS} holds, the correlation between $U_{e3}$ and
$\mu \to e + \gamma$ arises naturally.

Note that the relation of Eq. (\ref{PMNS}) is valid at the high scale and
thus it is important
to evaluate all the entries of the RH-side of this equation at the
high scale from their known values at the weak scale. Generically,
given that neutrino running effects are small for hierarchical neutrino
spectra, the running effects on the PMNS matrix appearing above are
neglected. While this is true for the other two of the angles in the
PMNS matrix, any small correction to $U_{e3}$ can have significant
impact on the value of the radiatively generated $(\Delta_{LL})_{12}$
entry in the slepton mass matrix.

As is well known, the form of the Yukawa matrix feeds into the flavour
violating  $LL$ entries of the slepton mass matrix through the well
known RG effects. At the leading log level, this expression is given by :
\begin{equation}
(\Delta_{LL})_{i\neq j} =
- \frac{3m^2_0 + A^2_0}{16\pi^2} \sum_k Y_{\nu \: ik}
Y_{\nu \: kj}^\dagger
\ln \left(\frac{M^2_X}{M^2_{R_k}} \right)
\label{deltaLL}
\end{equation}
where $m_0$ and $A_0$ are the common soft scalar mass and trilinear coupling,
$M_{R_k}$ the right-handed neutrinos masses and $M_X$ the energy scale at which the
SUSY breaking terms appear (coincident, in our framework, with the $SO(10)$
breaking scale). We have used the notation for the slepton mass matrices
as
\begin{equation}
\mathcal{M}_{\tilde{l}}^2 = \left( \begin{array}{cc}
\Delta_{LL} & \Delta_{LR} \\
\Delta_{RL} & \Delta_{RR} \end{array} \right),
\label{sleptonmatrix}
\end{equation}
where all the entries on the RHS are matrices in flavour space.
Expanding Eq.(\ref{deltaLL}) using the neutrino Yukawas of Eq. (\ref{PMNS}) we
have for the $12$ or equivalently $e\mu$ entry :
\begin{equation}
(\Delta_{LL})_{12} =
- \frac{3m^2_0 + A^2_0}{16\pi^2} \left( y_t^2 \; U_{e3} U_{\mu 3}^*
\ln \left(\frac{M^2_X}{M^2_{R_3}} \right) + y_c^2 \; U_{e2} U_{\mu 2}^*
\ln \left(\frac{M^2_X}{M^2_{R_2}} \right) + y_u^2 \; U_{e1} U_{\mu 1}^*
\ln \left(\frac{M^2_X}{M^2_{R_1}} \right)     \right)
\label{delta12}
\end{equation}
The $U_{e3}$ dependence of $(\Delta_{LL})_{12}$ is clear from the above
equation. Importantly, as we see from above, $U_{e3}$ couples with the
dominant contribution $\propto y_t^2 \sim \mathcal{O}(1)$. As a consequence,
a vanishing value of $U_{e3}$ would strongly suppress the flavour violating
mass insertion (MI) by a factor $y_c^2/y_t^2\sim 10^{-4}$. For small values of $U_{e3}$,
the term with top quark contribution would begin to dominate, once $U_{e3}$
crosses the limit value :
\begin{eqnarray}
|U_{e3}^{lim}|\approx \frac{y_c^2}{y_t^2} \; \frac{|U_{e2}|\cdot|U_{\mu 2}|}
{|U_{\mu 3}|} \; \frac{\ln M_X -\ln M_{R_2}}{\ln M_X -\ln M_{R_3}} \sim
\mathcal{O}(10^{-5}),
\label{ue3limit}
\end{eqnarray}
where we have taken $M_X$ to be of the order $10^{17}$ GeV. Here and
throughout the paper, the best fit values \cite{schwetz} for the
neutrino oscillations parameters were used ($\Delta m^2_{\rm sol}=
7.9\cdot 10^{-5}\, {\rm eV}^2$, $\Delta m^2_{\rm atm}= 2.6\cdot
10^{-3}\, {\rm eV}^2$, $\sin^2\theta_{12}=0.3$,
$\sin^2\theta_{23}=0.5$), apart from $U_{e3}$ that is considered as
a free parameter, as mentioned above. Further, for illustrative
purposes, here and later, we will be considering the limit where the
low scale value of $U_{e3}~\rightarrow 0$. The right-handed
neutrinos masses $M_{R_k}$ were obtained by solving the seesaw
equation (this is possible without uncertainties thanks to the
ansatz on the form of $Y_{\nu}$). The values we found for the $M_R$
eigenvalues are:

\be M_{R_1} = 3.4 \cdot 10^6 \mathrm{GeV}; \,\,\,\, M_{R_2} = 2.9
\cdot 10^{10} \mathrm{GeV}; \,\,\,\, M_{R_3} = 1.7 \cdot 10^{14}
\mathrm{GeV} \ee

The above analysis assumes a leading log approximation, where the RHS of
the Eqs. (\ref{delta12}, \ref{ue3limit}) are typically assumed to be constant
and taken to be their weak scale values, whereas the full running, would take
into consideration the running effects of the neutrino mixing parameters
also appearing on the RHS of these equations.
The important parameter here
is $U_{e3}$ which could be very small at the weak scale and could attain a
non-negligible value at the high scale.
To trace the $U_{e3}$ evolution, we can use  the following effective operator:
\begin{equation}
m_{\nu}(\mu) = Y_\nu(\mu) M^{-1}_R(\mu) Y^T_\nu(\mu)
\label{mnu}
\end{equation}
The RGEs for (\ref{mnu}) are given in the literature \cite{lindner,chankowski}.
From them, it is possible to estimate the generated $U_{e3}$ at high energy.
For instance, in the case of hierarchical neutrino spectrum ($m_{\nu_1} \ll m_{\nu_2}
\ll m_{\nu_3}$) and barring the PMNS phases, one gets\footnote{We will consider
all parameters to be real and set phases to be zero in the present work.
However
there are some subtleties associated with such an assumption especially in the
limit $U_{e3}$ goes to zero, which we will elaborate in the text.}:
\begin{eqnarray}
\Delta U_{e3}^{hie}(M_{W} \to M_{X}) &\approx&
-\frac{1}{16 \pi^2}
\left[ y^2_{\tau} \ln ( \frac{M_{X}}{M_{W}} )
+ y^2_{t} \ln ( \frac{M_{X}}{M_{R_3}} ) \right]
U_{e1} U_{e2} U_{\mu 3} U_{\tau 3}
\frac{m_{\nu_2} - m_{\nu_1}}{m_{\nu_3}}\nonumber\\
&\sim& - (\tan^2\beta \cdot \mathcal{O}(10^{-6}) +\mathcal{O}( 10^{-3})),
\label{deltaue3}
\end{eqnarray}
where the first contribution $\propto y_\tau^2 $ comes from the
ordinary MSSM RG corrections, whereas the second one $\propto y_t^2$ is
from neutrino Yukawa couplings above the seesaw scale.
\begin{figure}[t]
\psfrag{sps2}[c]{\huge{$\tan\beta=10$}}
\psfrag{ue3run}[c]{\huge{$U_{e3}(\mu)$}}
\psfrag{ue3}[c]{\huge{$U_{e3}(M_Z)$}}
\psfrag{MX}[r]{\Large{$\mu = M_X$}}
\psfrag{MU}[r]{\Large{$\mu = M_{GUT}$}}
\psfrag{MR}[r]{\Large{$\mu = M_{R_3}$}}
\includegraphics[angle=-90, width=0.58\textwidth]{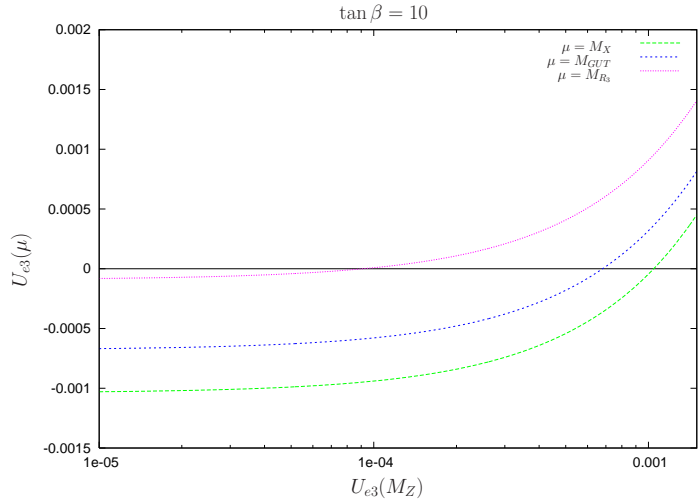}
\caption{Behaviour of high-scale values of $U_{e3}$ for
small $U_{e3}(M_Z)$. The neutrino spectrum is hierarchical and $\tan\beta = 10$}
\label{ue3run}
\end{figure}
Thus even when $U_{e3}\ll 10^{-3}$ at the weak scale, there is a generated
$\Delta U_{e3} \ger 10^{-3}$ at the high scale, especially at the scale where
it feeds into the slepton mass matrix.
The RG-generated $U_{e3}$ at the high scale
would now become the dominant contribution to the LFV as long as
the high-scale value of $U_{e3}$ overwhelms the limit value given
by Eq. (\ref{ue3limit}). For $\tan\beta \sim 10$ we see that, even
without the top-quark-like contribution, $\Delta U_{e3}$
far exceeds the limit value of $U_{e3}$ (\ref{ue3limit}). This RG
generated contribution is \textit{independent} of low-energy value
of $U_{e3}$ and would get generated at the high scale even when
$U_{e3}$ is zero. For larger values of $U_{e3}$ this contribution would
add to the low-energy number. This is best illustrated in the
Fig. \ref{ue3run}, where we plot the high-scale
value of $U_{e3}$ for a given value of $U_{e3}$ at the weak scale
\footnote{For the rest of the neutrino parameters required for the running,
we take $m_1 = 0.001$ eV, $\Delta m^2_{atm} > 0$.} at
three different high scales, $M_{R_3},~M_{GUT}$ and $M_X$. As we see
from the figure, $U_{e3}$ at the high scale takes a constant value
below $\sim \mathcal{O}(10^{-3}-~10^{-4})$ as a resultant of the RG
correction as per Eq. (\ref{deltaue3}).

\begin{figure}[t]
\psfrag{DeltaLL_12}[l]{\huge{$|\Delta^{LL}_{12}|, \, \mathrm{GeV}^2$}}
\psfrag{ue3}[c]{\huge{$U_{e3}(M_Z)$}}
\psfrag{title_mssm}[c]{\huge{{\bf SPS 2}}}
\includegraphics[angle=-90, width=0.58\textwidth]{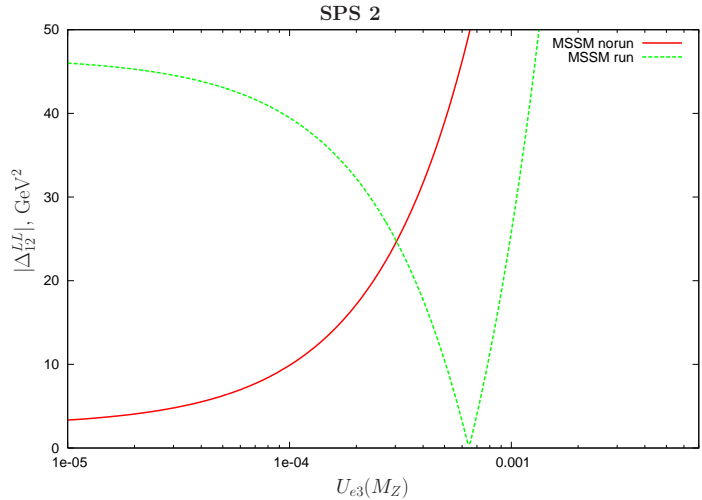}
\caption{Behaviour of $\Delta^{LL}_{12}$ for small values of $U_{e3}(M_Z)$,
in the point {\bf SPS 2} ($m_0 = 1450$ GeV, $M_{1/2} = 300$ GeV, $A_0 = 0$,
$\tan\beta = 10$) of the mSUGRA parameter space, with and without
$U_{e3}$ evolution.}
\label{deltarun}
\end{figure}

It is interesting to compare the contributions
in Eqs. (\ref{ue3limit}) and (\ref{deltaue3}). It is obvious that
for reasonable values of tan $\beta$,
even without the consideration of the contribution $\propto ~y_t^2$,
the RG generated $U_{e3}$ is always larger than the limit value
$U_{e3}^{lim}$ irrespective of the value of the $U_{e3}$ at the weak
scale, even if it is zero. This implies $(\Delta_{LL})_{12}$ will
always have a constant contribution due to the RG generated $U_{e3}$.
This is best demonstrated in Fig. \ref{deltarun} where we show the
contrast in the variation of the $(\Delta_{LL})_{12}$ with respect to
$U_{e3}$ when neutrino running effects are taken into consideration
or neglected. As we see from the figure, even with small or vanishing
values of $U_{e3}$ at the weak scale, $(\Delta_{LL})_{12}$ has a constant
value which is much larger than the value of $(\Delta_{LL})_{12}$ without
taking into consideration the running effects.

The evaluation of the contribution related to the running of $U_{e3}$
involves a subtlety which manifests itself in the above figure.
This subtlety arises because of the
evolution of $U_{e3}$ and the unknown CKM-like phase, $\delta$ of
the $U_{\mathrm{PMNS}}$ matrix\footnote{In standard notation
$U_{e3} \equiv \sin\theta_{13} e^{i \delta}$. Here we use $U_{e3}$
and $\sin\theta_{13}$ interchangeably, since we set the phases to zero.}.
Note that in the limit $U_{e3}$ goes to zero, $\delta$ remains undefined.
And the RGE for the $\delta$ diverges \cite{lindner}. In the present work,
as we scan $U_{e3}$ for very small values starting from zero, $U_{e3}$
takes values which are negative at the high scale. Given that we have set
all the phases to be zero in our work, this would correspond to the phase $\delta$
assuming a value $\pi$ at the high scale, if we insist on the standard
CKM-like parameterisation for the $U_{\mathrm{PMNS}}$ matrix to be valid also
at the high scale, where all the angles are defined to be in the first
quadrant ($0~<~\theta_{13}<~\pi/2$). Thus $\delta$ jumps from zero to
$\pi$ after the inclusion of RGE corrections\footnote{For a nice discussion
of this point, see \cite{lindner}.}. Hence the contributions to $(\Delta_{LL})_{12}$
proportional to $y_t^2$ and $y_c^2$ in Eq. (\ref{delta12}) have opposite signs.
As a consequence cancellations occur between the two contributions
at the high scale as $U_{e3}$ is varied. The exact cancellation occurs
when the high scale $U_{e3}$ value takes the $U_{e3}^{lim}$ (\ref{ue3limit}).
A dip occurs in the $(\Delta_{LL})_{12}$ when this happens as seen in
Fig. \ref{deltarun} and correspondingly in the branching ratio\footnote{
For the impact of Majorana phases in BR($\mu \to e + \gamma$), please
see \cite{petcov}.}.

\begin{figure}[t]
\psfrag{DeltaLL_12}[l]{\huge{$|\Delta^{LL}_{12}|$}}
\psfrag{x}[c]{\huge{$M_{1/2}$}}
\psfrag{y}[c]{\huge{$BR(\mu \to e + \gamma)\cdot 10^{11}$}}
\psfrag{M12}[c]{\huge{$M_{1/2}$}}
\psfrag{m0}[c]{\huge{$m_0$}}
\psfrag{mssm40}[c]{\huge{{MSSM, $\tan\beta = 40$}}}
\psfrag{mssm_40}[c]{\huge{{MSSM, $\tan\beta = 40$}}}
\psfrag{1E-11}[c]{\textcolor{blue}{\Large{$BR < 10^{-11} $}}}
\psfrag{1E-12}[c]{\textcolor{blue}{\Large{$BR < 10^{-12} $}}}
\psfrag{1E-13}[c]{\textcolor{blue}{\Large{$BR < 10^{-13} $}}}
\psfrag{1E-15}[c]{\textcolor{red}{\Large{$BR < 10^{-15} $}}}
\psfrag{1E-16}[c]{\textcolor{red}{\Large{$BR < 10^{-16} $}}}
\psfrag{1E-17}[c]{\textcolor{red}{\Large{$BR < 10^{-17} $}}}
\includegraphics[angle=-90, width=0.48\textwidth]{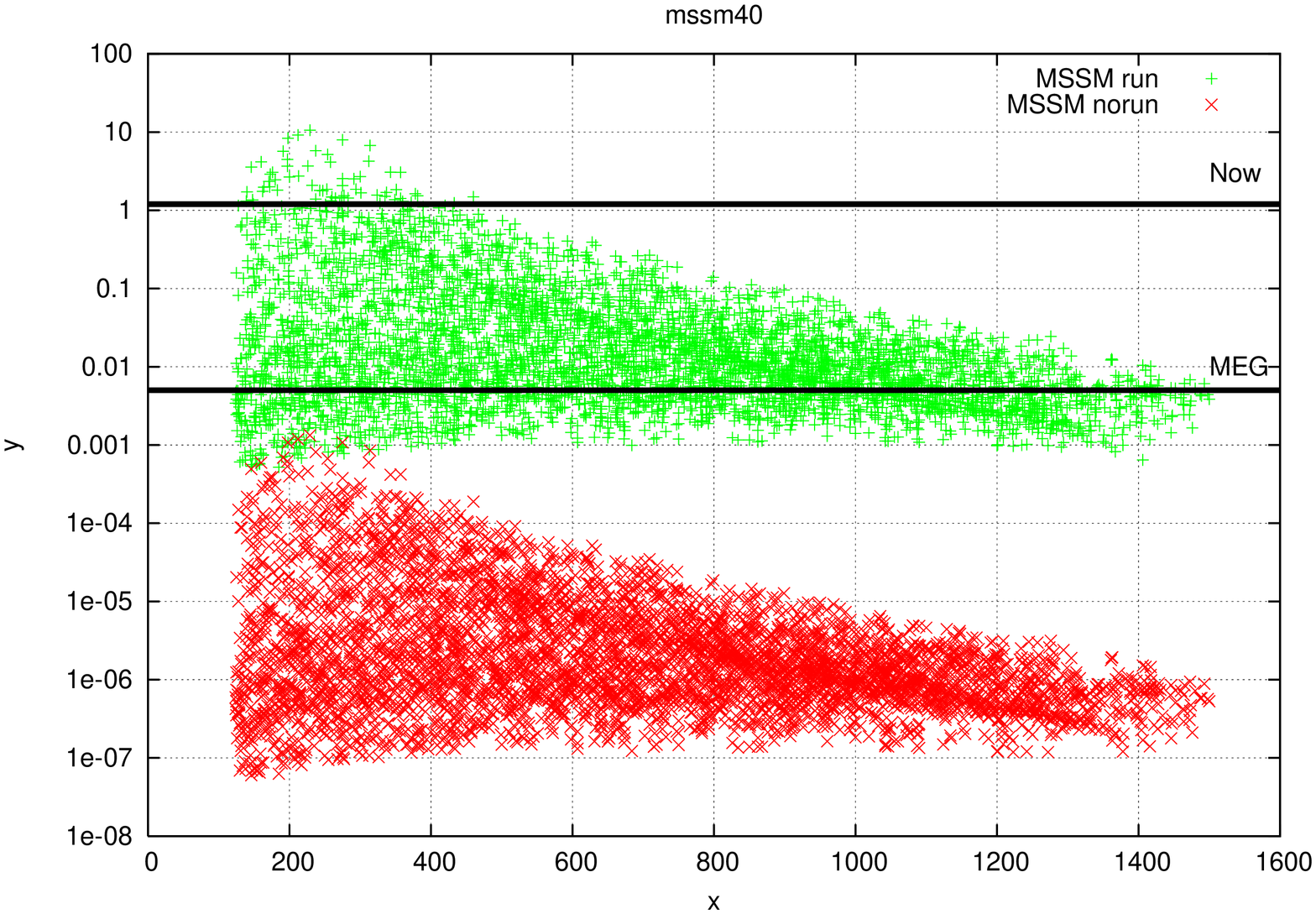}
\includegraphics[angle=-90, width=0.48\textwidth]{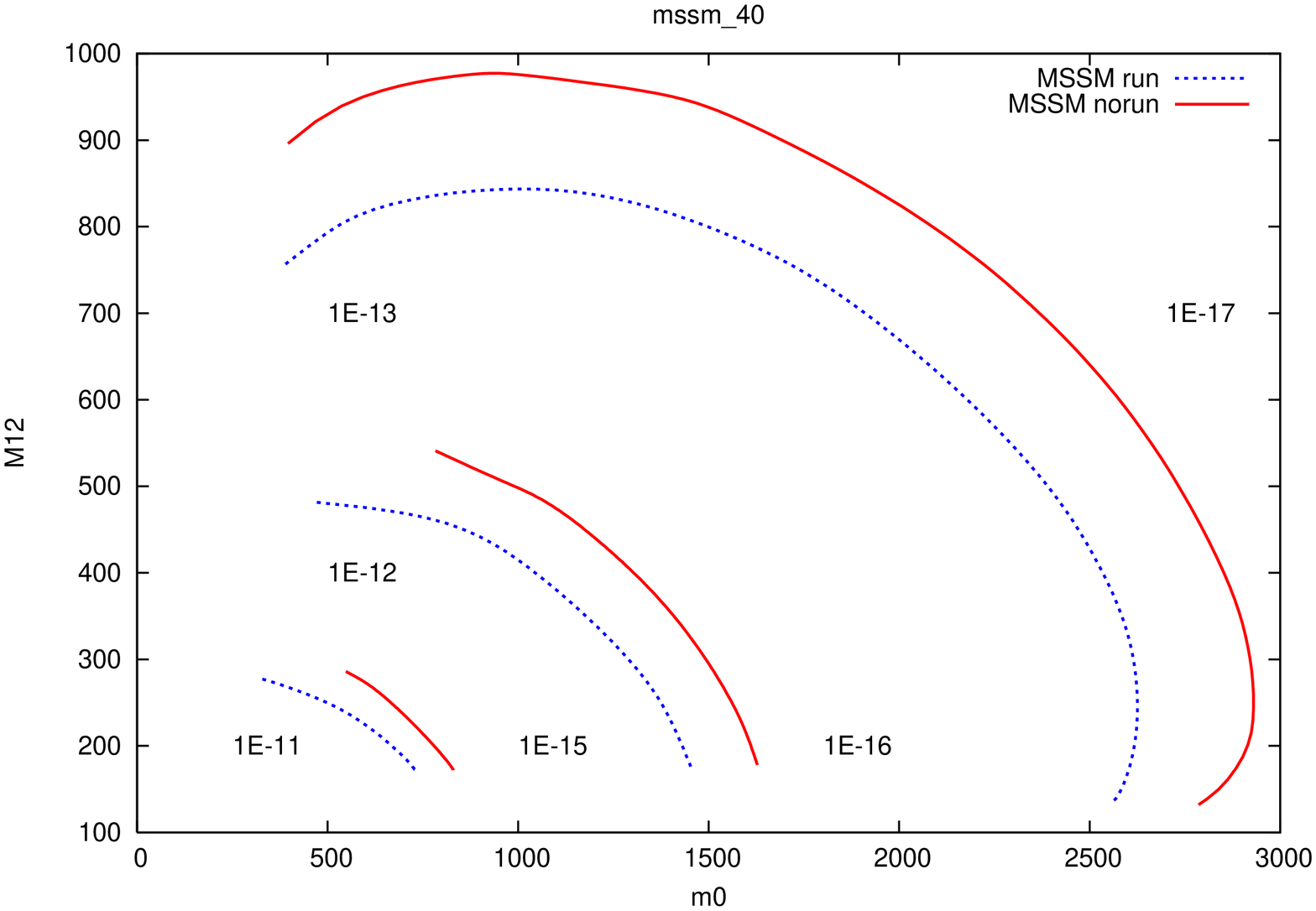}
\caption{BR($\mu \to e +\gamma$) scatter and contour plot
for $U_{e3}(M_Z)=0$ ($\tan\beta = 40$, $\mu > 0$), as a comparison between
the \textit{running} and
\textit{non-running} $U_{e3}$ cases. The scatter plot is made by scanning the
mSUGRA parameters ($0 < m_0 < 5 \mathrm{TeV}$,
$0 < M_{1/2} < 1.5 \mathrm{TeV}$, $-3 m_0 < A_0 < 3 m_0$)
and keeping the points within an approximate LHC accessible region
(i.e. $m_{\tilde{t}} \le 2.5 \mathrm{TeV}$).
For the contour plot $A_0 =0$. }
\label{scatcontour}
\end{figure}

Finally, we demonstrate the effect of taking into account the
RG corrections to the neutrino mass matrix on the scatter/exclusion
plots of SUSY-RNMSSM parameter space in Figs. \ref{scatcontour}.
We can see from the scatter plot that branching ratio increases by
at least three orders of magnitude once running effects are taken
into account, which can be traced to $\Delta U_{e3}$ which is
an order or two larger than the $U_{e3}^{lim}$. For large tan $\beta$
this increase is very significant: it brings most of the parameter
space into the realm of MEG experiment (this is apparent in the scatter
plot of Fig. \ref{scatcontour} where tan$\beta=40$).
In terms of the exclusion plots
in the $(m_0,M_{1/2})$ plane, we see that taking into consideration the
running effects largely enhances the region of parameter space probed
indirectly by the MEG experiment. While the above two plots exhibit
the particularly sizable effect of including the running of $U_{e3}$ when
tan$\beta$ is large, it should be noted that these effects
are always present and would be significant for whatever
value of tan$\beta$.

Finally before we close this section, a few comments are in order.
In the above analysis, we have parameterized the unknown $U_{e3}$ by
considering it as a free input parameter. We have further considered the
limit where it tends to zero at the low scale to illustrate the effect
of small values of $U_{e3}$ on the Br($\mu\to e,\gamma)$. This has been
done with a purely phenomenological perspective, without dealing with
the possible flavour symmetries giving rise to the present experimentally
determined form of the $U_{\mathrm{PMNS}}$ matrix.

The limit $U_{e3}(M_Z)\rightarrow0$ which we had discussed in this section,
however needs further clarifications in this respect\footnote{Incidentally,
let's note that the current best fit value is $U_{e3}=0$ 
\cite{schwetz}.}. In the limit
$U_{e3}\rightarrow0$, the neutrino mass matrix can be thought of as
having an additional (possibly discrete) symmetry \cite{Grimus:2004cc}\footnote{This
statement is typically defined in the basis where charged lepton mass matrix is diagonal.}.
Unless this symmetry
is broken, a non-zero value for $U_{e3}$ cannot be achieved. Let's note
that the simplest symmetries like the
$\mu \leftrightarrow \tau$ \cite{mohapatra} are
not really compatible with the $SO(10)$ seesaw framework and Yukawa
identification we have considered here\footnote{The required
$\mu \leftrightarrow\tau$ symmetric structure
of the $Y_{\nu}$ matrix cannot, for instance, be satisfied in the case of the
$Y_{\nu}$ eigenvalues having the same hierarchy of the $Y_{u}$.}.
In case there exists a flavour
symmetry which does lead to $U_{e3}(M_X) = 0$, then one can assume
that there exists some flavon fields breaking this symmetry, leading
to a nonzero value of $U_{e3}$ already at high scale.
Below such breaking scale the flavour symmetry is no more effective and
the RGEs are exactly as given in \cite{lindner,chankowski}.
If the flavour symmetry breaking also has a radiative origin, then effects
could be similar in magnitude with RG effects.
Thus, one can imagine partial cancellations between
these two effects leading to small values of $U_{e3}$ at the weak
scale, without requiring a large fine tuning (as in the case
$U_{e3}(M_Z)~\sim~10^{-4}$). On the other hand, $U_{e3}$ itself
can also be purely of radiative origin.
Finally we note here what we need is only small value of
$U_{e3}$: $0<U_{e3}(M_Z)<10^{-3}$, as in our case $U_{e3}$
decreases from $M_X$ to $M_Z$ (due to absence of phases, the
RG running carries the same sign at all stages), with the limit
value $U_{e3}\rightarrow0$ only used to emphasize this case.

\section{Running $U_{e3}$ and Double flavour violating MI in GuTs}
So far most of the discussion in the previous sections has been
focused on the impact of taking into consideration the RG effects
of neutrino mass matrix within the context of RNMSSM. There we have
assumed a $U_{e3}^{lim}$ value which is similar in the context of
SUSY-GUTs. However in SUSY-GUTs, additional RG running  effects
between $M_X$ and $M_{GUT}$
exist which can become dominant in some regions of the parameter space.
\begin{figure}[t]
\psfrag{A}[c]{$(\Delta_{LL})_{23}$}
\psfrag{B}[c]{$m_{\tau} \mu \tan \beta$}
\psfrag{C}[c]{$(\Delta_{RR})_{31}$}
\psfrag{D}[c]{$\mu_L$}
\psfrag{E}[c]{$e_R$}
\psfrag{F}[c]{$M_1$}
\psfrag{G}[c]{$\tilde{\mu}_L$}
\psfrag{H}[c]{$\tilde{\tau}_L$}
\psfrag{I}[c]{$\tilde{\tau}_R$}
\psfrag{L}[c]{$\tilde{e}_R$}
\psfrag{M}[c]{$\tilde{B}$}
\psfrag{N}[c]{$\tilde{B}$}
\psfrag{O}[c]{$\gamma$}
\includegraphics[angle=0, width=0.58\textwidth]{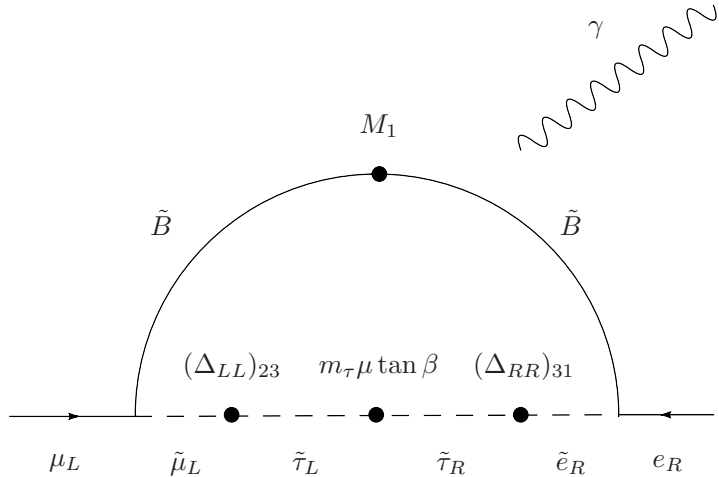}
\caption{Feynman diagram contributing to the double MI of Eq. \eref{doubleMI}.}
\label{doublefeyn}
\end{figure}

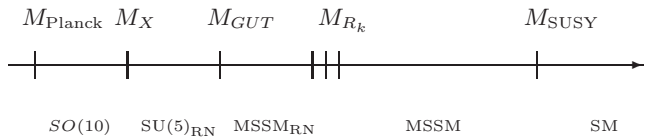
\begin{figure}[h]
{\footnotesize
\begin{center}
\begin{picture}(280,65)

\put(0,40){\vector(1,0){240}}

\put(5,55){$M_{\mathrm{Planck}}$}
\put(10,45){\line(0,-1){10}}
\put(40,55){$M_X$}
\put(45,45){\line(0,-1){10}}
\put(75,55){$M_{GUT}$}
\put(80,45){\line(0,-1){10}}

\put(117,55){$M_{R_k}$}
\put(115,45){\line(0,-1){10}}
\put(120,45){\line(0,-1){10}}
\put(125,45){\line(0,-1){10}}

\put(195,55){$M_{\mathrm{SUSY}}$}
\put(200,45){\line(0,-1){10}}

\put(15,15){\tiny $SO(10)$}
\put(50,15){\tiny  $\mathrm{SU(5)}_{\mathrm{RN}}$}
\put(85,15){\tiny $\mathrm{MSSM}_{\mathrm{RN}}$}
\put(150,15){\tiny MSSM}
\put(220,15){\tiny SM}
\end{picture}
\end{center}
}
\caption{\label{scales}Schematic
 picture of the energy scales involved in the model.}
\end{figure}

In terms of the so-called mass-insertion approximation \cite{MIA},
the main such contribution is a double mass-insertion which generates
an `effective' LR flavour violating mass entry \cite{calibbi}:
\begin{equation}
(\delta_{LR})^{eff}_{21} =
(\delta_{LL})_{23} \cdot \mu m_\tau \tan\beta \cdot (\delta_{RR})_{31}
\label{doubleMI}
\end{equation}
where the $\delta_{ij} \equiv \Delta_{ij}/m_{\tilde{l}}^2$; where
$\Delta_{ij}$ is already defined in Eq. (\ref{sleptonmatrix}) of the
slepton mass matrix and $m_{\tilde{l}}^2$ is the average slepton
mass.
The origin of such double mass-insertion is best depicted
in the Feynman diagram in Fig. \ref{doublefeyn} \cite{paradisi}.
This contribution, which is independent of $U_{e3}$ would provide a
flat contribution to the branching ratio irrespective of the value of
$U_{e3}$ at the weak scale.
To discuss in detail, we will work in the $SO(10)$
framework, where $SO(10)$ is broken down to the Standard Model through
an intermediate scale of SU(5) located at around $10^{16}$ GeV. The
various scales involved here can be summarised in the Fig. \ref{scales}.

\begin{figure}[t]
\psfrag{mssm}[c]{\huge{MSSM contour plot}}
\psfrag{su5}[c]{\huge{SU(5) contour plot}}
\psfrag{m0}[c]{\huge{$m_0$}}
\psfrag{M12}[c]{\huge{$M_{1/2}$}}
\psfrag{sps2}[l]{\Large{\bf SPS 2}}
\psfrag{sps3}[l]{\Large{\bf SPS 3}}
\psfrag{I}[c]{\huge{{\bf I}}}
\psfrag{II}[c]{\huge{{\bf II}}}
\psfrag{1E-13}[c]{\huge{$BR < 10^{-13} $}}
\psfrag{1E-14}[c]{\huge{$BR < 10^{-14} $}}
\psfrag{1E-15}[c]{\huge{$BR < 10^{-15} $}}
\includegraphics[angle=-90, width=0.58\textwidth]{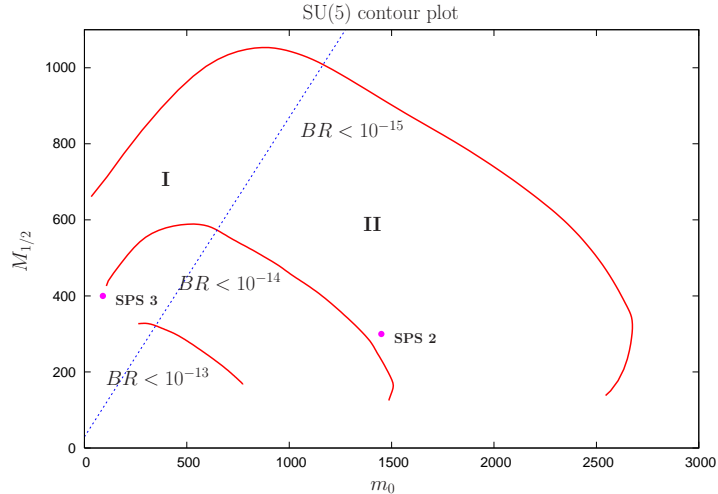}
\caption{$U_{e3}(M_Z)=0$ contour plots at fixed BR($\mu \to e +\gamma$) with $A_0=0$,
$\tan\beta =10$ for SU(5) ($\Delta m_{atm} > 0$, $m_1 = 10^{-3}
\mathrm{eV}$). Two regions of the $(m_0,M_{1/2})$ plane are
defined: I where the dominant contribution to BR($\mu \to e +\gamma$) is given
by $(\delta_{LL})_{23}(\delta_{RR})_{13}$; II where $(\delta_{LL})_{12}$
dominates. }
\label{contour10}
\end{figure}

Given the impact of the $U_{e3}$ running in SUSY-seesaw (Sect. II), one
would expect that the $U_{e3}$ proportional contribution would be
the dominant force within the SUSY-GUT
framework as the neutrino mass matrix running effects are larger.
However, the double flavour violating MI Eq. (\ref{doubleMI}) which is
independent $U_{e3}$ could become dominant in some regions of the parameter
space. In these regions the running of $U_{e3}$ would have no \textit{strong
impact} on the total branching ratio. The interplay between these two
effects is best demonstrated in Fig. \ref{contour10}\footnote{In making
 this figure, we have taken, $M_{GUT} \sim 2\cdot 10^{16} \mathrm{GeV}$,
$M_X \sim 5\cdot 10^{17} \mathrm{GeV}$.
The numerical routine computes the rates of LFV processes
by using the exact masses and mixings of the SUSY particles, obtained from full
1-loop RGE evolution of the mSUGRA parameters. The high-energy  values of
fermion masses and mixings are set by evolving them from the e.w. scale up
to $M_X$.  For more details about the numerical routine,
 we refer to \cite{calibbi}.}.

Fig. \ref{contour10} shows the contour plots at fixed BR($\mu \to e +\gamma$)
in the plane $(m_0,M_{1/2})$ of the mSUGRA parameter space, for $U_{e3}=0$,
normal neutrino hierarchy and lightest neutrino mass $m_1 = 10^{-3}
\mathrm{eV}$; the other mSUGRA parameters are set to
be: $A_0=0$, $|\mu|>0$ and $\tan \beta = 10$.
The diagonal line in the center of the figure are the points where the
contribution of the double insertion
$(\Delta_{LL})_{23}(\Delta_{RR})_{13}$ is equal to the
$(\Delta_{LL})_{12}$. This line divides into two regions the
$(m_0,M_{1/2})$ plane: region I where the double insertion dominates, and
region II where $(\Delta_{LL})_{12}$ forms the main contribution.
Thus though generically, $(\Delta_{LL})_{12}$ dominates the amplitudes for
$\mu \to e + \gamma$, for extremely small values of $U_{e3}$,
the contributions of $(\Delta_{LL})_{12}$, enhanced by the running of $U_{e3}$,
and the double MI can be competing in some regions of the SUSY parameter space.

\begin{figure}[t]
\psfrag{y}[c]{\huge{$BR(\mu \to e +\gamma)\cdot 10^{11}$}}
\psfrag{x}[c]{\huge{$U_{e3}(M_Z)$}}
\psfrag{full}[r]{\Large{full computation}}
\psfrag{d12}[r]{\large{$\delta_{12}^{LL}$ only}}
\psfrag{d12rr}[r]{\large{$\delta_{12}^{RR}$ only}}
\psfrag{di}[r]{\large{$\delta_{23}^{LL}\delta_{13}^{RR}$ only}}
\psfrag{sps2}[c]{\huge{{\bf SPS 2}, SU(5)}}
\psfrag{sps3}[c]{\huge{{\bf SPS 3}, SU(5)}}
\includegraphics[angle=-90, width=0.48\textwidth]{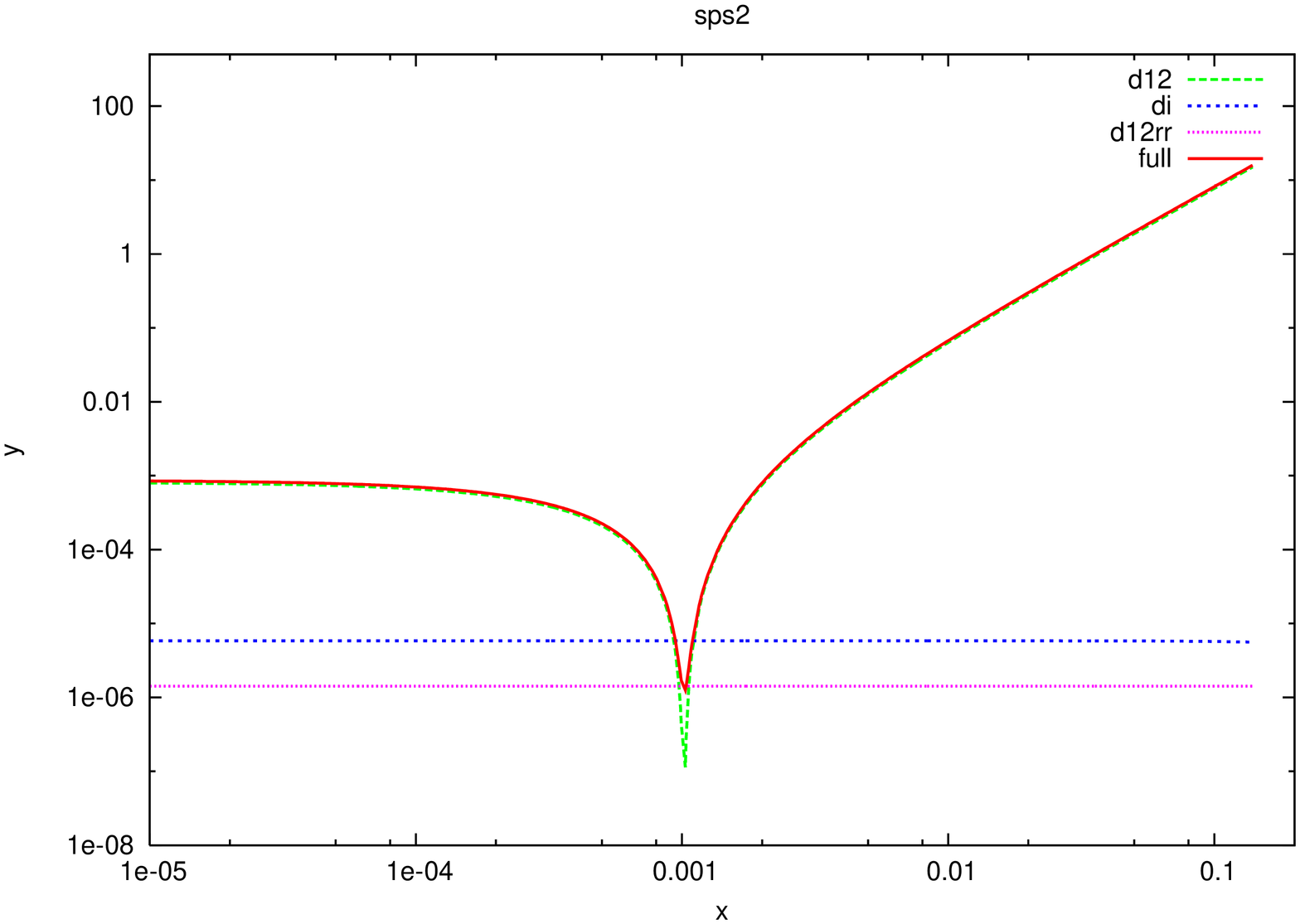}
\includegraphics[angle=-90, width=0.48\textwidth]{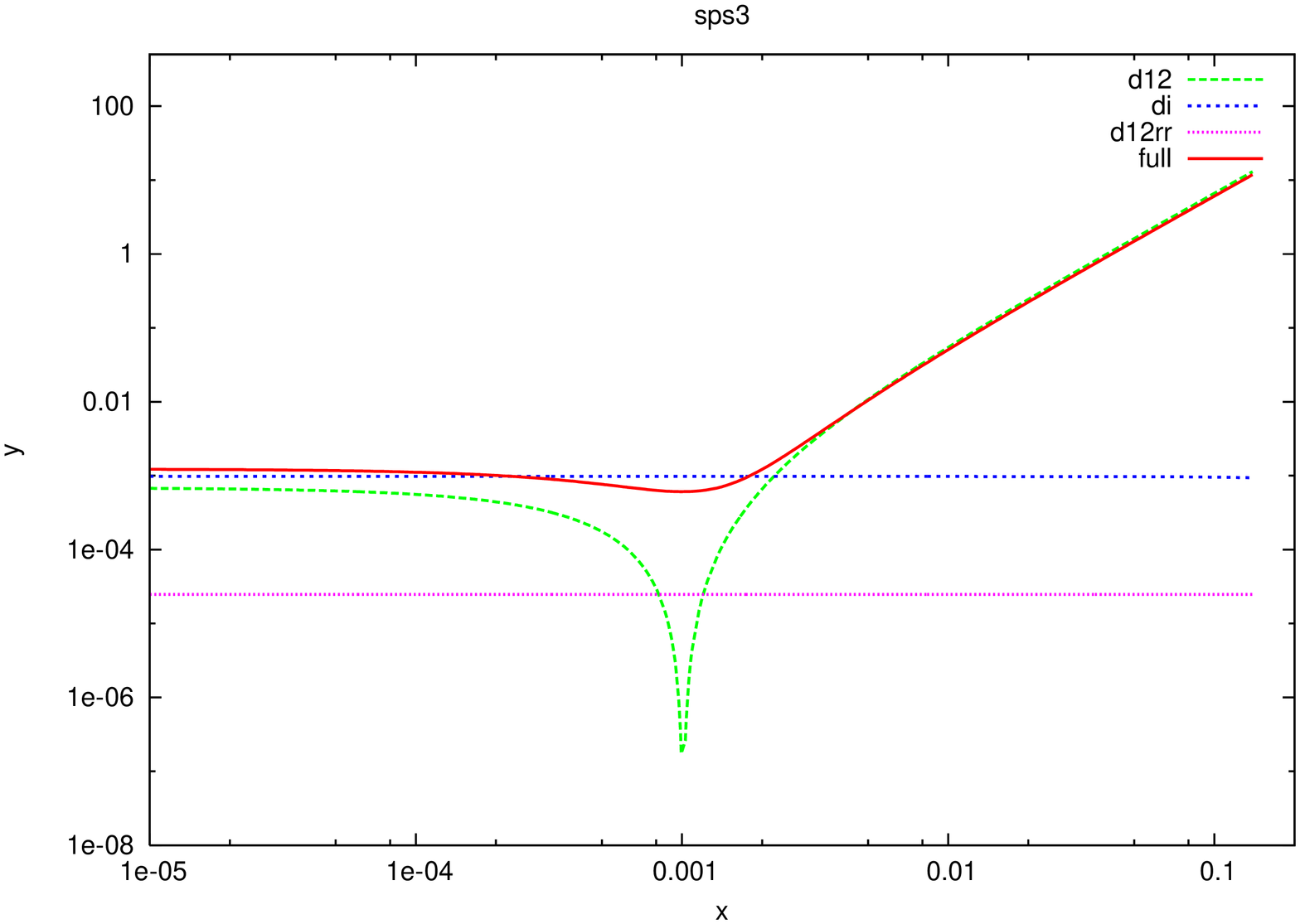}
\caption{Different SU(5) contributions to BR($\mu \to e +\gamma$)
for the benchmark points, as a function of low-energy $U_{e3}$.}
\label{deltaplot}
\end{figure}

The benchmark points {\bf SPS 3} ($m_0 = 90$ GeV, $M_{1/2} = 400$ GeV,
$A_0 = 0$, $\tan\beta = 10$) and {\bf SPS 2} ($m_0 = 1450$ GeV,
$M_{1/2} = 300$ GeV, $A_0 = 0$, $\tan\beta = 10$) lie respectively
in region I and in region II. The competition between these two
contributions is evident in Fig. \ref{deltaplot}, where
BR($\mu \to e +\gamma$) is plotted as a function of
$U_{e3}(M_Z)$ considering only one contribution at a time (i.e. putting
the other ones to zero) in the two different regions I ({\bf SPS 3})
and II ({\bf SPS 2}). As we can see from the figures, in both the
cases, above $\sim~ 10^{(-3)}$, the $(\Delta_{LL})_{12}$ contribution dominates.
However, below that value, in the case of {\bf SPS 2}, the running effects of
$U_{e3}$ become very crucial, whereas in the case of {\bf SPS 3}, the double
flavour violating mass insertion dominates. Finally we note that
the $U_{e3}$ dependent minima in the $(\Delta_{LL})_{12}$ contribution are
due to the cancellation of two dominant contributions as discussed
in the previous section.

\begin{figure}[t]
\psfrag{y}[c]{\huge{$BR(\mu \to e +\gamma)\cdot 10^{11}$}}
\psfrag{x}[c]{\huge{$U_{e3}(M_Z)$}}
\psfrag{sps2}[c]{\huge{{\bf SPS 2}}}
\psfrag{sps3}[c]{\huge{{\bf SPS 3}}}
\includegraphics[angle=-90, width=0.48\textwidth]{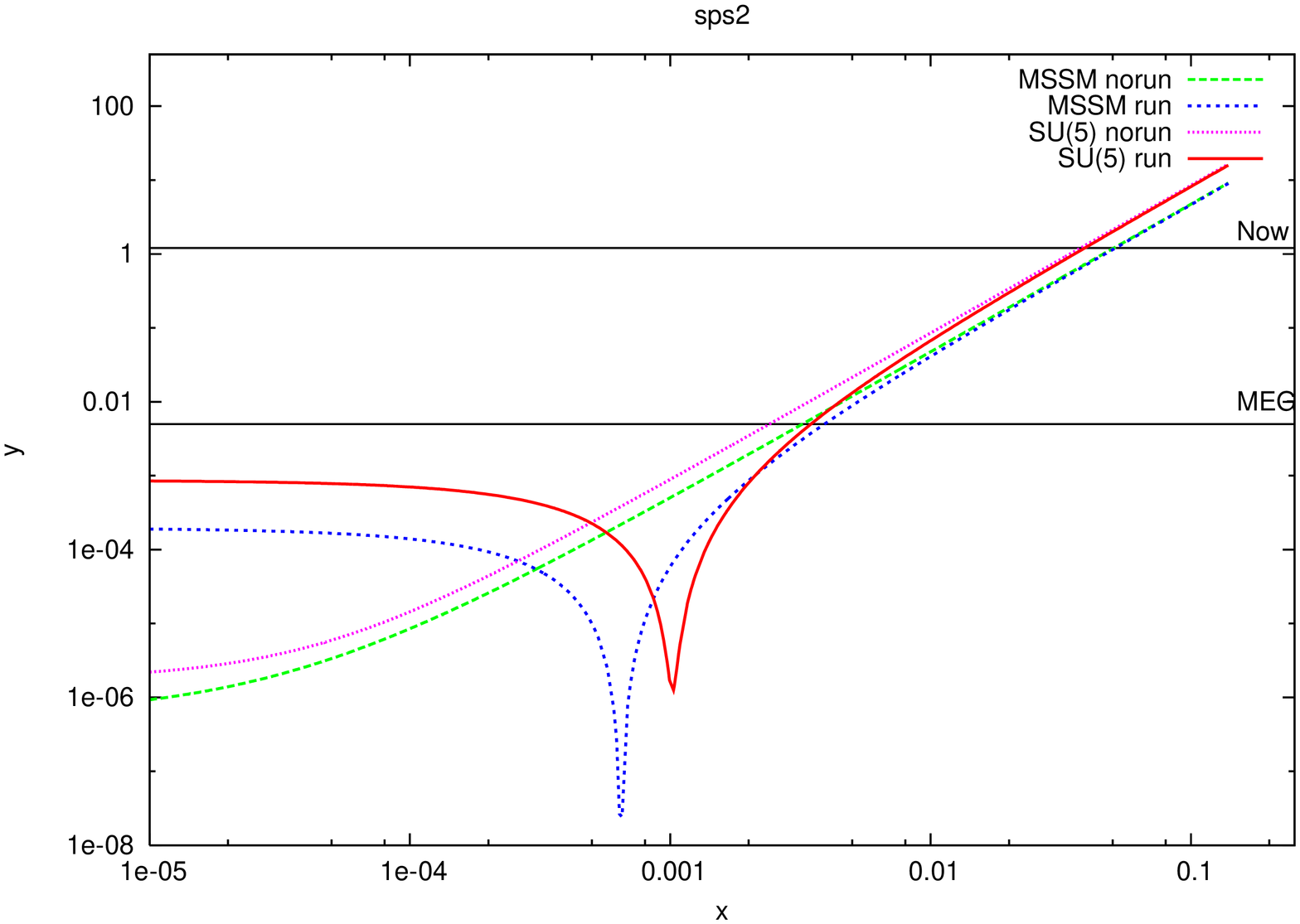}
\includegraphics[angle=-90, width=0.48\textwidth]{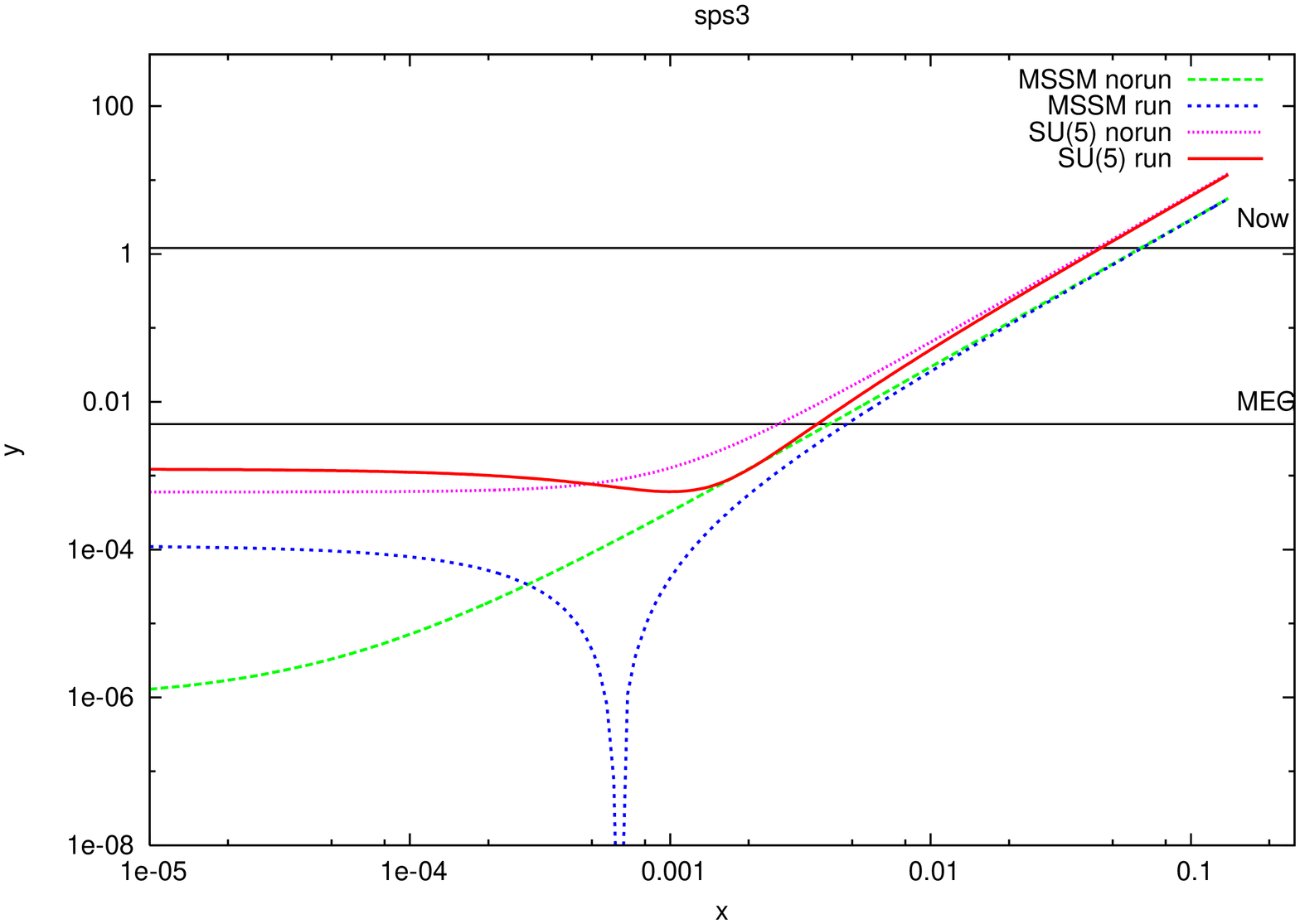}
\caption{ BR($\mu \to e +\gamma$) as a function of $U_{e3}(M_Z)$
in the {\bf SPS 2}, {\bf SPS 3} benchmark points of the mSUGRA
 parameter space ($\Delta m_{atm} > 0$, $m_1 = 10^{-3} \mathrm{eV}$).
The plots show the effect of switching on the $U_{e3}$ running
 both for MSSM and SU(5).}
\label{plot10}
\end{figure}

It would be interesting to compare the effect of running in both the
case of GUT theories and in RNMSSM. In Fig. \ref{plot10}, we plot
the total branching ratios with and without taking the running effects
for case of RNMSSM and SUSY-SU(5).  In RNMSSM the $U_{e3}$ running gives,
for small low-energy values, an order of magnitude enhancement of the BR
with respect to the sub-dominant $y_c$ contribution. In SU(5) such enhancement
is almost hidden by pure SU(5) effect in the case of {\bf SPS 3} that lies
in region I, while it results dominant effects for {\bf SPS 2}.

So far we have been discussing  the RG effects of $U_{e3}$ within
the context of hierarchical neutrino spectrum.  Different light neutrinos
spectra could consistently change the above results.
While in the case of degenerate spectrum (lightest neutrino mass
$\ger 0.1 \mathrm{eV}$), we found similar results to the
normal hierarchy (even if this the degenerate case should be more sensitive
to the change of phases). The so-called inverted hierarchy ($\Delta m_{atm} < 0$,
$m_{\nu_3} = 10^{-3} \mathrm{eV}$) doesn't give enhancement effects
due to the $U_{e3}$ running comparable to the normal hierarchy case.
This is as expected from the direct proportionality of $\Delta U_{e3}$ to
the lightest neutrino mass $m_{\nu_3}$ \cite{lindner}. Moreover in this latter
case, the scales of right handed neutrinos are much closer to the GUT scale and thus
even the pure SU(5) effects coming from double MI which can enhance
the BR over the $y_c$ contribution are smaller compared to the normal hierarchical
case.

\begin{figure}[t]
\psfrag{y}[c]{\huge{$BR(\mu \to e +\gamma)\cdot 10^{11}$}}
\psfrag{x}[c]{\huge{$U_{e3}(M_Z)$}}
\psfrag{su5_40}[c]{\huge{SU(5),  $\tan\beta=40$}}
\psfrag{sps2_40}[c]{\huge{{\bf SPS 2}, $\tan\beta=40$}}
\psfrag{m0}[c]{\huge{$m_0$}}
\psfrag{M12}[c]{\huge{$M_{1/2}$}}
\psfrag{1E-11}[c]{\huge{$BR < 10^{-11} $}}
\psfrag{1E-12}[c]{\huge{$BR < 10^{-12} $}}
\psfrag{1E-13}[c]{\huge{$BR < 10^{-13} $}}
\includegraphics[angle=-90, width=0.48\textwidth]{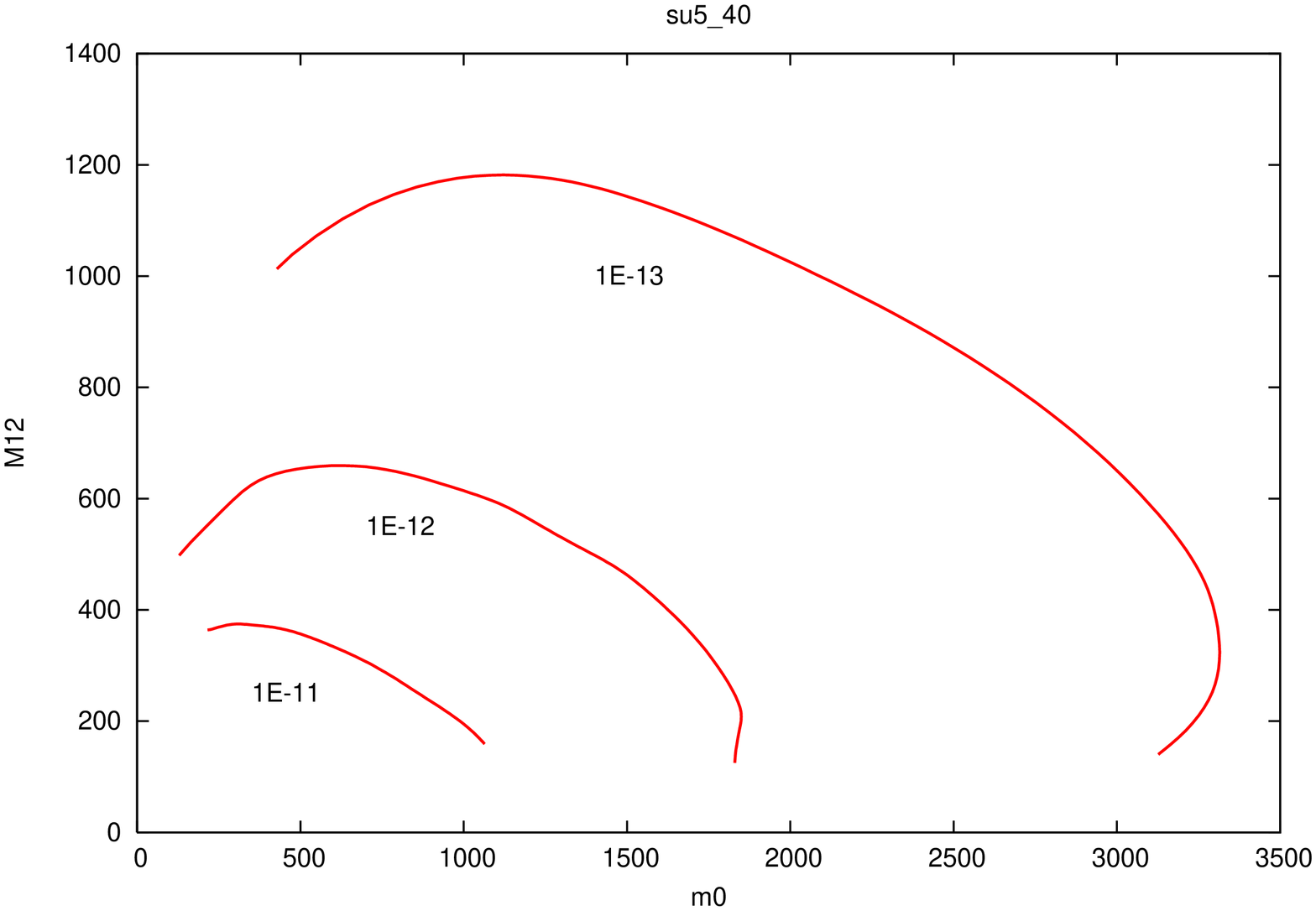}
\includegraphics[angle=-90, width=0.48\textwidth]{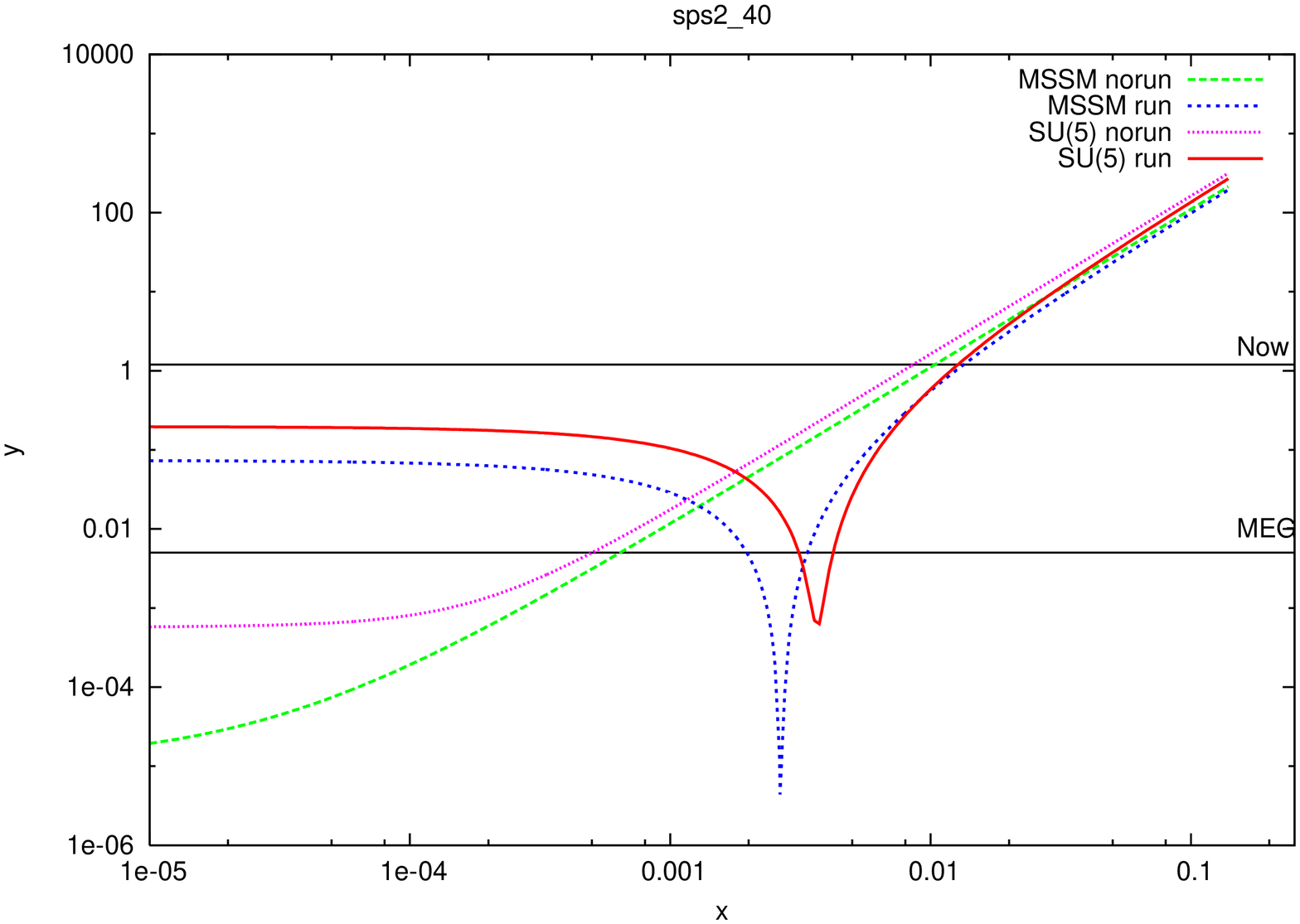}
\caption{$U_{e3}(M_Z)=0$ contour plot and
BR($\mu \to e +\gamma$) vs. $U_{e3}(M_Z)$ for
$\tan\beta = 40$. The point considered for the second plot is
{\bf SPS 2} with $\tan\beta = 40$ instead of 10.  }
\label{plot40}
\end{figure}

Finally, we consider what happens in the case of $\tan\beta = 40$.
We find that for small values of $U_{e3}$ the dependence of the branching
ratio on $\tan\beta$, where $(\delta_{LL})_{12}$ gives the dominant
contribution, is not the usual $\propto (\tan\beta)^2$, because
$\tan\beta$ would also affect the running of $U_{e3}$
(and $(\delta_{LL})_{12}\propto U_{e3}$). The result is that the enhancement
of branching ratio at $\tan\beta = 40$ with respect to  $\tan\beta = 10$
is much larger than the usual scaling factor of 16. This can be seen
in Fig. \ref{plot40}, where the SU(5) contour plot and BR($\mu \to e +\gamma$)
for {\bf SPS 2} with $\tan\beta = 40$ instead of 10 are plotted.

\section{Final Remarks}
The unknown neutrino mixing angle $U_{e3}$ is an object of much
speculation and interest for neutrino mass model builders as well
as for experimentalists. While we have no clue on the value of this
angle, except for an upper bound of $|U_{e3}|~ \ler 0.14$, it could
as well take very small values even reaching zero at the weak scale.
Over the past few years, various strategies have been devised
to probe $U_{e3}$ down to values of $\mathcal{O}(10^{-2})$
\cite{ue3expts}. While these experiments probe $U_{e3}$ at weak
scale values, SUSY-seesaw based models give information about the
high-scale values of $U_{e3}$ through indirect measurements of decay
rates such as $\mu \to e + \gamma$ at dedicated facilities like
MEG \cite{megexpt}. In the present paper, we have stressed the
importance of considering RG running effects on the neutrino mass
matrices while making such a correlation between weak scale measurements
and high-scale probes of $U_{e3}$, which has been neglected in
earlier works \cite{neutrinoposter}.

Let's make a final consideration on the link between the value of $U_{e3}$
(at the low-energy scale at which we hope to measure it soon) and
BR($\mu\to e +\gamma$). The key-point is the assumption that the angles
entering the diagonalization of the neutrino Yukawa matrix $Y_{\nu}$ are
linked to those connected with the diagonalization of the mass matrix
of physical neutrinos, i.e. the PMNS angles. If this is the case, then,
independently of what we assume for the value of the $Y_{\nu}$ eigenvalues,
it is possible that the running of $U_{e3}$ from the electroweak scale
up to $M_X$ induces an effect on it of $\mathcal{O}(\tan^2\beta \cdot 10^{-6})$.
Hence, even if future measurements of $U_{e3}$ would lead to very small values
of it, such running effects could provide contributions to $U_{e3}$ so large
that a BR($\mu\to e +\gamma$) accessible to the MEG experiment could occur
if some large neutrino coupling is present. This is the main point of our
present analysis where a top-down approach with an underlying $SO(10)$
symmetry is taken. An analogous investigation
\cite{herrero} where a bottom-up phenomenological approach was considered
also had similar conclusions.
The main difference between the two approaches concerns the information
about the size of the neutrino Yukawa couplings. In our $SO(10)$ framework, we
can correlate $Y_{\nu}$ with $Y_u$, hence obtaining the large contribution
from the running from $M_R$ up to $M_X$ in Eq. (\ref{delta12}).
Also, assuming such a GUT underlying structure allowed us to include the
effects due to the running above $M_{GUT}$ of Eq. (\ref{doubleMI}).

In conclusion, let us stress again that taking
running effects into account could in principle lead to a ``constant'' enhancement
of the value of $U_{e3}$ at the high scale, bringing $\mu \to e + \gamma$
into the realm of MEG for SUSY parameter space regions which were
previously excluded without the inclusion of such running.

We have not addressed the important and interesting issue of origins
of neutrino mixing angles particularly $U_{e3}$, treating it as a
free parameter. The question of low values of $U_{e3}$ at the weak
scale, when the radiative corrections themselves are as large as
$\sim 10^{-3}$ would deserve more attention as it points out to
cancellations within the parameters of the model and radiative corrections.
We hope to deal with this issue at a later date.

The correlation of $U_{e3}$ and flavour violating effects continue to be
important in the context of SUSY-GUTs and any measurement of flavour violation
at MEG could lead to shedding some light on either $U_{e3}$ or on the parameter
space of SUSY-GUTs.

\textbf{Acknowledgements}
SKV acknowledges support from Indo-French Centre for Promotion of
Advanced Research (CEFIPRA) project No:  2904-2 `Brane World
Phenomenology'.  He is also partially supported by INTAS grant,
03-51-6346, CNRS PICS \# 2530, RTN contract MRTN-CT-2004-005104
and by a European Union Excellence Grant, MEXT-CT-2003-509661.
LC, AF and AM thank the PRIN `Astroparticle Physics' of the Italian
Ministry MIUR and the INFN `Astroparticle Physics' special project.
We also aknowledge support from the RTN European program MRTN-CT-2004-503369
 `The Quest for Unification' and MRTN-CT-2006-035863 `UniverseNet'.
LC thanks the Ecole Polytechnique--CPHT for hospitality.
LC, AM and SKV also thank the CERN Theory Group for hospitality
during various stages of this work. We thank S. Antusch,
M. J. Herrero and  A. Teixeira for bringing to our notice their
work which is related to the topic discussed here.
In particular, AM acknowledges an interesting discussion
with M. J. Herrero. LC thanks P. Paradisi and O. Vives for useful
comments and suggestions.

\end{document}